  \documentclass[conference,a4paper]{IEEEtran}
  \IEEEoverridecommandlockouts
  \usepackage{cite}
  \usepackage{amsmath,amssymb,amsfonts}
  \usepackage{algorithmic}
  \usepackage{graphicx,epstopdf}
  \usepackage{textcomp}
  \usepackage{xcolor}
  \usepackage{setspace}
  \usepackage{enumitem}
  \def\BibTeX{{\rm B\kern-.05em{\sc i\kern-.025em b}\kern-.08em
  		T\kern-.1667em\lower.7ex\hbox{E}\kern-.125emX}}
  \usepackage{subfigure}
  \newtheorem{my_theorem}{Theorem}

  \newtheorem{my_proposition}{Proposition}
  
  \DeclareMathOperator*{\argmin}{arg\,min}
\title{ Terahertz Wireless Transmissions with  Maximal Ratio Combining over  Fluctuating Two-Ray Fading}
\author{
	\IEEEauthorblockN{Atharva Anand Joshi, Pranay Bhardwaj, and S. M. Zafaruddin}\\
	\IEEEauthorblockA{ Department of Electrical and Electronics Engineering, 
		BITS Pilani, Pilani Campus, Pilani-333031, Rajasthan, India\\ Email: \{f20180515, p20200026, syed.zafaruddin\}@pilani.bits-pilani.ac.in}
	
	\thanks{This work was supported in part by the Start-up Research Grant, Department of Science Technology (DST), Science and Engineering Research Board (SERB), India under Start-up Research Grant SRG/2019/002345.}
}

\begin{document}
	\maketitle
	
	\begin{abstract}		
Mitigating channel fading and transceiver impairments are desirable for high-speed terahertz (THz) wireless links.  This paper analyzes the performance of a multi-antenna  THz wireless system by considering the combined effect of pointing errors and fluctuating two-ray (FTR) fading model. We provide a statistical characterization of the maximal ratio combining (MRC) receiver  over independent and nonidentical (i.ni.d.) channel conditions in terms of multi-variate Fox's H  by deriving density and  distribution functions of the signal-to-noise ratio (SNR) of a single-link THz link using incomplete Gamma function.  We develop exact analytical expressions of  outage probability,  average bit-error-rate (BER), and ergodic capacity for both single-antenna  and MRC receivers. We also present the diversity order of the system by deriving asymptotic expressions for outage probability and average BER at high SNR to obtain insights into  the system performance. We validate our derived analytical expressions with   Monte-Carlo simulations and demonstrate the effect of various system and channel parameters on the performance of single and multi-antenna THz wireless communications.

	\end{abstract}
	
\section{Introduction}\label{sec:introduction}
Terahertz (THz) wireless is an upcoming technology to provide  new spectrum resources for future communication systems.  The availability of  contiguous high bandwidth transmissions in  the THz spectrum can be potential  for wireless backhaul/fronthaul technology \cite{Koenig_2013_nature,Wang2014, Elayan_2019}. The THz spectrum is mostly unlicensed and can support secured terabits per second (Tbps) data transmissions  with low  latency  for various high-end applications. The  line-of-sight (LOS) THz technology requires high directional antennas with higher gain to compensate for severe path loss due to the molecular absorption of transmitted signals.  Nevertheless, the THz link is  susceptible to the  random pointing errors caused by the misalignment between transmitter and receiver antenna beams and may incur transceiver distortion at higher frequencies in addition to the stochastic multi-path fading \cite{Kokkoniemi_2018,KOKKONIEMI2020, Boulogeorgos_Analytical, Shanyun2021}. Alleviating the adverse effects of signal attenuation and fading is desirable  for high-speed THz links.

Recently,  dual-hop and multi-hop relaying at THz frequencies have  been investigated  \cite{ Xia_2017,Giorgos2020, Boulogeorgos_2020_THz_THz,huang2021, Boulogeorgos_Error,Pranay_2021_TVT, Rong_2017, Abbasi_2017,Mir2020}. More specifically, the authors in \cite{Xia_2017} formulated an optimal relaying distance  for THz-band communication to maximize the network throughput.  In \cite{Giorgos2020},  a relay selection approach was suggested to mitigate the  impact of  antenna misalignment and shadowing due to the human blockage in a multi-relay setup.  A reconfigurable intelligent surface (RIS) assisted   multi-hop THz system over Rician fading was considered in \cite{huang2021} to mitigate the signal attenuation using deep reinforcement learning (DRL) based beam-forming technique. Considering the generalized  $\alpha$-$\mu$ fading combined with stochastic  pointing errors,  the decode-and-forward (DF) protocol was employed to link THz and radio frequency (RF) technologies \cite{Boulogeorgos_Error,Pranay_2021_TVT}. Using multi-antenna transceivers, the DF relaying was studied for  a dual-hop  THz-THz link \cite{Boulogeorgos_2020_THz_THz}.

There has been an increased research interest to model the short-term fading for THz communications
 \cite{Riza2017, Kursat2021, Papasotiriou2021,RIS_THz_HW_Impaiment}. A Gamma mixture channel model for THz transmissions for a short ($<1$ \mbox{m}) link is proposed in \cite{Kursat2021}. The authors in  \cite{Papasotiriou2021} find $\alpha$-$\mu$ fading model suitable for  the THz transmission using the measurement  at $152$\mbox{GHz} for a link length within
  $50$\mbox{m}. Using  a comprehensive THz measurement data at $300$\mbox{GHz} for train-to-infrastructure and inside station \cite{Ke2019}, the authors in \cite {RIS_THz_HW_Impaiment} demonstrate that the fluctuating two ray (FTR) model is a better fit for THz multi-path fading modeling than the conventional Rician and Nakagami-m distributions. Using the combined effect of the  short-term FTR fading,   antenna   misalignment,  and hardware impairments, \cite {RIS_THz_HW_Impaiment} derived outage probability and ergodic   capacity for  RIS-aided THz   systems. 
  
 The recently proposed FTR model has been extensively studied for mmWave wireless transmissions \cite{Juan2017, Jiayi2018, Hui2019, Wen2018,Zhang_2020_mmWave_FSO,Osmah2019}.
In \cite{Juan2017,Jiayi2018, Hui2019}, analytical performance was studied for a single-link FTR fading channel.   The physical layer secrecy performance over the FTR fading channel was analyzed \cite{Wen2018}.  The authors in \cite{Zhang_2020_mmWave_FSO} analyzed  the performance of a mixed free-space-optics (FSO)-mmWave system by modeling the mmWave and FSO channels as FTR and Gamma-Gamma distributed, respectively.  In contrast to the single-antenna system,  multi-antenna at the receiver can harness the spatial diversity over independent fading for improved performance  \cite{Hahemi2020, Hussein2021, Jiakang2019,Maryam1019, Hadi2020}.  In \cite{Hahemi2020}, the authors analyzed the performance of an  equal gain combining (EGC) receiver by deriving outage probability and average BER using single-variate mathematical functions over FTR fading channels.  In \cite{Hussein2021}, a low complexity selection combining receiver was investigated with a performance analysis on the outage probability, average BER, and ergodic capacity in terms of multi-variate Fox's H function.     In \cite{Jiakang2019}, the authors analyzed the optimal maximal ratio combining (MRC)  receiver by deriving the PDF and CDF of the sum of arbitrarily distributed FTR variates. In \cite{Maryam1019}, the outage probability and an upper bound on average BER was derived for the MRC receiver. The authors in   \cite{Hadi2020} provided   asymptotic and non-asymptotic expressions of the outage probability and average BER for the MRC over  non-identical distributed FTR fading channels. The moment matching method was used to approximate  statistics of the sum of FTR fading channel for  analyzing  relay-assisted radio frequency (RF)-mmWave wireless communications for   high-speed trains \cite{Jiayi2020}.

In the light of THE above research and to the best of the author's knowledge, performance analysis of the MRC receiver with THz wireless transmissions over FTR fading channels jointly with stochastic pointing errors is not available in the open literature.   Our main contributions of this paper are  follows:
\begin{itemize}[leftmargin=*]
	\item By  deriving PDF and CDF of a single-link THz link using standard mathematical functions, we provide exact  statistical characterization of the SNR for the MRC receiver under the  joint effect of  FTR short-term fading  and zero-bore sight pointing errors considering independent and 
	nonidentical (i.ni.d.) channel conditions in terms of multi-variate Fox's H function. 
	\item We develop  exact analytical expressions of  ergodic capacity, outage probability, and average BER  for both single-antenna reception  and MRC receiver and present asymptotic expressions  for outage probability and average BER at high SNR. We derive the diversity order of the considered system to  show the advantage of multi-antenna reception and   the impact of pointing errors.
	\item We evaluate multi-variate Fox's H function using Python code  \cite{Alhennawi2016} and validate our derived analytical expressions with   Monte-Carlo simulations. We also demonstrate the effect of various system  and channel parameters on the performance of THz wireless communications.
\end{itemize}

\section{System Model}
 A single-antenna source communicates to an $L$-antenna destination over the THz spectrum. The THz link is affected by path-loss, short-term fading, pointing errors, and transceiver distortions. Assuming negligible hardware impairments \cite{Boulogeorgos_Error, Pranay_2021_VTC}, the received signal $y_i$ at the $i$-th antenna is given by:
\begin{equation} y_{i} = h_l h_i s + w_i, 
\label{eq:rx_one}
\end{equation}
where $h_{l}$ is the path gain,  $s$ is the transmitted signal with power $P$, $h_i$ denotes the fading channel coefficient, and $w_i$ is the additive white Gaussian noise with a variance $\sigma_w^2$. 
 The deterministic path gain $h_{l}$ is dependent on antenna gains, frequency, and  molecular absorption coefficient \cite{Boulogeorgos_Analytical}:	
\begin{equation}
h_l = \frac{c\sqrt{G_{t}G_{r}}}{4\pi f d} \exp(-\frac{1}{2}k(f)d)
\end{equation}
where $c$, $f$, and $d$ respectively denote the speed of light,	carrier frequency,  link distance whereas $G_{t}$ and $G_{r}$ denote  gain of the  transmitting antenna and receiving antenna, respectively. The term  $k(f,T,\psi,p)$ is the molecular absorption coefficient  depends on the temperature $T$, relative humidity $\psi$, and atmospheric pressure $p$ \cite{Boulogeorgos_performance_2018}.

The compound channel coefficient is $h_i = h_{pi} h_{fi}$, where  $h_{pi}$ and $h_{fi}$ models  pointing error  and short term fading, respectively. We use the zero boresight model  for pointing errors $h_{pi}$  \cite{Farid2007}:
\begin{equation}
\begin{aligned}
f_{h_{pi}}(h_p) &= \frac{\phi^2}{S_{0}^{\phi^2}}h_{p}^{\phi^{2}-1},0 \leq h_p \leq S_0,
\end{aligned}
\label{eq:pdf_hp}
\end{equation}
where $S_0=\mbox{erf}(\upsilon)^2$ with $\upsilon=\sqrt{\pi/2}\ (r_1/\omega_z)$ and $\omega_z$ is the beam-width, $\phi = {\frac{\omega_{z_{\rm eq}}}{2 \sigma_{s}}}$ with  $\omega_{z_{\rm eq}}$ as the equivalent beam-width at the receiver, which is given as $\omega_{z_{\rm eq}}^2 = {\omega^2_z} \sqrt{\pi} \mbox{erf}(\upsilon)/(2\upsilon\exp(-\upsilon^2)) $, and $\sigma^2_{s}$ is the variance of pointing errors displacement characterized by the horizontal sway and elevation \cite{Farid2007}.

To model $|h_{fi}|^2$, we use the  FTR fading channel with PDF given as \cite{Jiayi2018}:	 
\begin{eqnarray} \label{eq:ftr1}
&{f_{|h_{fi}|^{2} }}(x) = \frac{{{m^m}}}{{\Gamma (m)}}\sum \limits _{j = 0}^\infty {\frac{{K^jd_jx^j}}{{(\Gamma (j + 1))^2(2\sigma ^2)^{j+1}}}} \exp (- \frac{x}{2\sigma ^2})
\end{eqnarray}
where $K$ is the ratio of the average power of the dominant component  and  multi-path, $m$ is the index of fading severity, and $\Delta$ denotes the similarity   of two dominant waves. The term $\sigma^2$ represents the variance of diffused components such that $\sigma^2=\frac{1}{2(1+K)}$ for the normalized averaged SNR. The factor $d_j$  is defined in \cite{Jiayi2018} and  recently updated with an additional factor in \cite{Miguel2021}.

We denote  SNR   of  the $i$-th antenna  as 	$ \gamma_i=\gamma_0|h_i|^2$, where $\gamma_0= \frac{P |h_{l}|^2}{\sigma_{w}^2}$.
Assuming perfect channel state information (CSI),   SNR with optimal combining  for the MRC receiver is $\gamma_{}= \sum_{i=1}^L \gamma_i$. 	Thus, PDF and CDF  for the sum of product of pointing errors and FTR random variables is required for statistical performance analysis of the   MRC receiver.  

\section{Statistical Derivations}
In this section, we provide statistical results for the sum of $L$ arbitrarily distributed FTR fading combined with stochastic pointing errors by   deriving   closed-form expressions of the PDF and CDF of the single THz link.  
\begin{my_proposition}
If $|h_{i}|=|h_{fi}||h_{pi}|$ is the combined effect of FTR  fading and   pointing errors, then the PDF and CDF of the single-link  SNR $\gamma_i=\gamma_0|h_{i}|^{2}$   are given by
\begin{eqnarray} \label{eq:pdf_single}
&	f_{\gamma_i}(x) = \frac{{{\phi^{2}}{m^m}}}{{2S_0^{\phi^{2}} \gamma_0^{\frac{\phi^2}{2}} {{\left({2{\sigma ^2}} \right)}^{{\frac{\phi^{2}}{2}}}}} {\Gamma (m)}}\sum \limits _{j = 0}^\infty {\frac{{K^jd_j}}{{[\Gamma \left({j + 1} \right)]^2}}} \nonumber \\& {{x^{({\frac{\phi^2}{2}}-1)}}}\Gamma \left({ -\frac{\phi^{2}}{2}+j+1,\frac{{x}}{{2 \gamma_0 \sigma ^2 S_0^{2}}}} \right) 
\end{eqnarray}
\begin{eqnarray}\label{eq:cdf_single}
	&F_{\gamma_i}(x)= \frac{{{m^m}}}{{2S_0^{\phi^{2}}} \gamma_0^{\frac{\phi^2+1}{2}}{{\left({2{\sigma ^2}} \right)}^{{\frac{\phi^{2}}{2}}}}{\Gamma (m)}}\sum \limits _{j = 0}^\infty {\frac{{{K^j}{d_j}}}{{ [\Gamma ({j + 1}})]^2}} \nonumber \\& 2 x^{\frac{\phi^2}{2}} \Gamma({ -\frac{\phi^{2}}{2}+j+1,\frac{{x}}{{2\gamma_0 \sigma^2 S_0^{2}}}})\nonumber\\& -2^{{\phi^{2}}{}+1} {({}{{\gamma_0\sigma ^2 S_0^{2}}})}^{\frac{\phi^2}{2}} x^{-\frac{\phi^2}{2}} \Gamma (j+1,\frac{{x}}{{2\gamma_0\sigma ^2 S_0^{2}}})
\end{eqnarray}
\end{my_proposition}
\begin{IEEEproof}
Transforming random variable in \eqref{eq:pdf_hp}, we get 
\begin{eqnarray}\label{eq:pe1}
		 f_{{h_{pi}^{2}}}(x)=\frac{1}{2} {\phi ^{2}S^{-\phi ^{2}}_{0}}{}x^{{\frac{\phi ^{2}}{2}-1}}, \quad 0 \leq x \leq {S_{0}^{2}},
\end{eqnarray}	 
Using the limits  of PDF  in \eqref{eq:ftr1} and \eqref{eq:pe1}, the PDF of 	$|h_{i}|^{2}= h_{fi}h_{p}^{2}$	can be expressed as \cite{papoulis_2002} 
	\begin{eqnarray}\label{eq:pdf_der1}
		 f_{h_{i}^{2}}(x) = \int _{0}^{S_{o}^2} \frac {1}{y} f_{h_{fi}^{}}\left ({\frac {x}{y}}\right) f_{h_{pi}^2}(y) \mathrm {d}y.
		  \end{eqnarray}
Substituting \eqref{eq:ftr1} and \eqref{eq:pe1} in \eqref{eq:pdf_der1}, we have
	\begin{eqnarray}\label{eq:pdf_der2}
	&	f_{h_{i}^{2}}(x)= \sum \limits _{j = 0}^\infty {{\frac{{{K^j}{d_j}}}{{[\Gamma(j+1)]^2}{(2{\sigma^{2}})^{j+1}}}}{x^{j}}}\nonumber \\& \int_{0}^{S_{0}^2} {e^{-\frac{x}{2y{\sigma ^2}}}} {y}^{\frac{\phi^{2}}{2}-2-j}dy 
	\end{eqnarray}
Expressing the integral in \eqref{eq:pdf_der2} in terms of incomplete Gamma function with a transformation of random variable $\frac{1}{\gamma_0}f_{|h_{fp}|^{2}}(\gamma/\gamma_0)$, we get  \eqref{eq:pdf_single}. To derive the CDF $F_{\gamma}(x)=\int_0^{x}	f_{\gamma}(z)dz$, we use the identity $\int x^{b-1} \Gamma(s, x) \mathrm{d} x= -\frac{1}{b}\big(x^b\Gamma(s,x)+\Gamma(s+b,x)\big)$  to get \eqref{eq:cdf_single}.
\end{IEEEproof}
In the following Theorem, we capitalize the results of Proposition 1 to derive PDF and CDF of the SNR for the MRC receiver:
\begin{my_theorem}
If the PDF of single THz link is distributed as \eqref{eq:pdf_single}, then PDF $f_{\gamma_{}}(\gamma)$ and CDF $F_{\gamma_{}}(\gamma)$ of the SNR $\gamma_{}= \sum_{i=1}^L \gamma_i$  for an  $L$-antenna MRC receiver are given by
	\begin{flalign}\label{eq:pdf_mrc}
		f_{\gamma_{}}(\gamma) =& \frac{1}{\gamma}\sum \limits _{{[j_i=0]}_{i=1}^L}^\infty \prod_{l=1}^{L} \frac{{{\phi^{2}}{{m_l}^{m_l}}}}{{2S_0^{\phi^{2}}} {\Gamma (m_l)}} {{\frac{{{{K_l}^{j_l}}{{d_l}_{j_l}}\gamma^{\phi^2/2}}}{{[\Gamma^{}(j_l+1)]^2}{(2{\sigma_l^{2}}\gamma_0)^{\phi^2/2}}}}} \nonumber \\ &
		~~~~~~H^{0,0:2,1;2,1;\dots;2,1}_{0,1:2,2;2,2;\dots;2,2}  \bigg[ ~\begin{matrix} V(\gamma) \end{matrix}~ \bigg| \begin{matrix} ~~V_1~~ \\ ~~V_2~~  \end{matrix} \bigg] 
	\end{flalign}
where $V(\gamma) = \big\{\frac{\gamma}{{2\sigma_i^2 S_0^{2} \gamma_0}}\big\}_{i=1}^L$,  $V_1 = -: \big\{ \big(1-\frac{\phi^2}{2},1 \big), \big(1,1\big) \big\};\cdots;\big\{ \big(1-\frac{\phi^2}{2},1 \big), \big(1,1\big) \big\} $ and $V_2 = \big\{ \big(1-\frac{L\phi^2}{2};1,\cdots,1 \big) \big\}; \big\{ \big(-\frac{\phi^2}{2}+j_i+1,1 \big), \big(0,1\big) \big\}_{i=1}^L$	
\begin{eqnarray}\label{eq:cdf_mrc}
	&	F_{\gamma_{}}(\gamma) = \sum \limits _{{[j_i=0]}_{i=1}^L}^\infty \prod_{l=1}^{L} \frac{{{\phi^{2}}{{m_l}^{m_l}}}}{{2S_0^{\phi^{2}}} {\Gamma (m_l)}} {{\frac{{{{K_l}^{j_l}}{{d_l}_{j_l}}\gamma^{\phi^2/2}}}{{[\Gamma^{}(j_l+1)]^2}{(2{\sigma_l^{2}\gamma_0})^{\phi^2/2}}}}}  \nonumber \\ & ~	H^{0,0:2,1;2,1;\dots;2,1}_{0,1:2,2;2,2;\dots;2,2}\bigg[~ \begin{matrix} U(\gamma) \end{matrix}~ \bigg| \begin{matrix} ~~U_1~~ \\ ~~U_2~~ \end{matrix} \bigg]
\end{eqnarray}
where $U(\gamma) = \big\{\frac{\gamma}{{2\sigma_i^2 S_0^{2} \gamma_0}}\big\}_{i=1}^L$, $U_1 = -: \big\{ \big(1-\frac{\phi^2}{2},1 \big), \big(1,1\big) \big\}; \cdots ;\big\{ \big(1-\frac{\phi^2}{2},1 \big), \big(1,1\big) \big\} $ and $U_2 = \big\{ \big(-\frac{L\phi^2}{2};1,\cdots,1 \big) \big\}; \big\{ \big(-\frac{\phi^2}{2}+j_i+1,1 \big), \big(0,1\big) \big\}_{i=1}^L$	
\begin{figure*}
	\begin{eqnarray}\label{eq:asymptotic_out}
		&P_{\rm out_{}}^{{\rm MRC}, \infty} = \hspace{0 mm}\sum \limits _{{[j_i=0]}_{i=1}^L}^\infty \hspace{0mm} \prod_{l=1}^{L} \frac{{{\phi^{2}}{{m_l}^{m_l}}}}{{2A_0^{\phi^{2}}} {\Gamma (m_l)}} {{\frac{{{{K_l}^{j_l}}{{d_l}_{j_l}}\gamma^{\phi^2/2}}}{{[\Gamma^{}(j_l+1)]^2}{(2{\sigma_l^{2}\gamma_0})^{\phi^2/2}}}}} \frac{1}{ \beta \Gamma\big(\hspace{0 mm}\frac{L\phi^2}{2} +\sum_{i=1}^{L} g_i\hspace{0mm}\big)} \nonumber \\&\prod_{i=1}^{L} \frac{\prod_{j=1,j\neq c_i}^{2} \Gamma(b_{i,j}+B_{i,j}-B_{i,j}g_i) \Gamma\big(\hspace{0mm}\frac{\phi^2}{2} - 1+ g_i\hspace{0mm}\big)  }{\Gamma\big(\frac{\phi^2}{2}-j_i\big) } \left(\hspace{0mm}\frac{\gamma}{{2\sigma_i^2 S_0^{2}\gamma_0}}\hspace{0mm}\right)^{g_i}
	\end{eqnarray}
	
	\begin{eqnarray}\label{eq:asymptotic_ber}
		&\bar{P}_{e}^{{\rm MRC}, \infty} = \frac{1}{2\Gamma(p)q^{\frac{L\phi^2}{2}}}\sum \limits _{{[j_i=0]}_{i=1}^L}^\infty \prod_{l=1}^{L} \frac{{{\phi^{2}}{{m_l}^{m_l}}}}{{2S_0^{\phi^{2}}} {\Gamma (m_l)}} {{\frac{{{{K_l}^{j_l}}{{d_l}_{j_l}}}}{{[\Gamma^{}(j_l+1)]^2}{(2{\sigma_l^{2}}\gamma_0)^{\phi^2/2}}}}}     \frac{\Gamma\big(p+\frac{L\phi^2}{2} +\sum_{i=1}^{L} g_i\big) }{ \beta \Gamma\big(\frac{L\phi^2}{2} + \sum_{i=1}^{L} g_i\big)} \nonumber \\ & \prod_{i=1}^{L} \frac{\prod_{j=1,j\neq c_i}^{2} \Gamma(b_{i,j}+B_{i,j}-B_{i,j}g_i) \Gamma\big(\frac{\phi^2}{2} - 1+ g_i\big) }{\Gamma\big(\frac{\phi^2}{2}-j_i\big) } \left(\frac{1}{{2\sigma_i^2 S_0^{2}\gamma_0}q} \right)^{g_i-1}
	\end{eqnarray}
where $b_{i,1} = -\frac{\phi^2}{2}+j+1$, $b_{i,2} = 0$, $B_{i,1} = 1$, $B_{i,2} = 1$  $g_i = \min\{-\frac{\phi^2}{2}+j_i+1, 0 \} $, $\beta = \prod_{i=1}^{L} B_{i,c_i}$, and $c_i = { \argmin}_{j=1:m_i} \left\{\frac{b_{i,j}}{B_{i,j}}\right\}$.
\hrule
\end{figure*}
\end{my_theorem}
\begin{IEEEproof}
	See Appendix A.
	\end{IEEEproof}
\section{Performance Analysis}
In this section, we analyze the performance of single-antenna THz link in terms of standard Mathematical functions and use multi-variate Fox's H  and Gamma functions to provide exact and asymptotic analysis on the  multi-antenna reception.
\subsection{Single Antenna Reception (SAR)}
\subsubsection{Outage Probability}
Outage probability is defined as the probability of instantaneous SNR  less than a predetermined threshold value $\gamma_{\rm th}$ i.e., $ P_{\rm out}^{} = P(\gamma <\gamma_{\rm th})$. Thus an exact expression of the outage probability  is $P_{\rm out}^{\rm SAR}= F_{\gamma_i}(\gamma_{\rm th})$, where $F_{\gamma_i}(x)$ is given in \eqref{eq:cdf_single}.
  \subsubsection{Average BER}
 Average BER is another important performance metric for communication systems,  defined as \cite{Ansari2011}
   \begin{equation} \label{eq:ber_eqn}
 \bar{P}_e = \frac{q^p}{2\Gamma(p)}\int_{0}^{\infty} \exp({-qx})x^{p-1}\Psi_{X}(x)dx
 \end{equation}
where $p$  and $q$ are constants to specify different modulation techniques and $\Psi_{X}$ is the CDF.  We denote by $\xi = {}{2\gamma_0\sigma^2S_0^2}$.  Using \eqref{eq:cdf_single} in \eqref{eq:ber_eqn}, applying the identity [\cite{Gradshteyn},6.455/1] and expressing the Hypergeometric function into Gamma function we get
\begin{eqnarray}\label{ber_single}
	&\bar{P}_{e}^{\rm SAR} = \frac{m^m p^q}{{4S_0^{\phi^{2}}} \gamma_0^{\frac{\phi^2+1}{2}}{{\left({2{\sigma ^2}} \right)}^{{\frac{\phi^{2}}{2}}}}\Gamma(m)\Gamma(p)}\sum \limits _{j = 0}^\infty \frac{{K^jd_j}}{[\Gamma(j+1)]^2} \nonumber \\ & \times \frac{2 (\xi)^{-(\frac{\phi^2}{2}+p)} e^{-q\gamma} \big[-2^{\frac{\phi^2}{2}} (\frac{\phi^2}{2}+p) \Gamma(j+1+p) + p \Gamma(2j+2+p) \big] }{p(\frac{\phi^2}{2}+p)}
\end{eqnarray}
 \subsubsection{Ergodic Capacity}
The ergodic capacity is defined as 
\begin{eqnarray}\label{eq:rate_eqn}
	\bar{\eta} = \int_{0}^{\infty} {\log_2}(1+x) \psi_X(x) dx
\end{eqnarray}
where $\psi_X$ denotes the PDF. Using \eqref{eq:pdf_single} in \eqref{eq:rate_eqn} and applying identity of definite integration of two Meijer's G function [\cite{Meijers},07.34.21.0011.01], we get an exact expression for the ergodic capacity of the single THz link 
\begin{eqnarray}\label{eq:rate_single_exact}
	&	\bar{\eta}^{\rm SAR} = \frac{{{m^m}}}{{\log (2)S_0^{\phi^{2}} \gamma_0^{\frac{\phi^2}{2}} {{\left({2{\sigma ^2}} \right)}^{{\frac{\phi^{2}}{2}}}}} {\Gamma (m)}}\sum \limits _{j = 0}^\infty \frac{{{K^j}{d_j}}}{[ \Gamma(j+1)]^2}  \xi^{-\frac{\phi^2}{2}} \nonumber \\& G_{3,4}^{4,1} \Bigg( \begin{matrix} -\frac{\phi^2}{2}, 1-\frac{\phi^2}{2}, 1 \\ -\frac{\phi^2}{2}+j+1, 0, -\frac{\phi^2}{2}, -\frac{\phi^2}{2}	\end{matrix} \Bigg| \xi     \Bigg)
\end{eqnarray}
Further,  we can use ${\rm log}(1+\gamma)>{\rm log}(\gamma)$ in \eqref{eq:rate_eqn} with \eqref{eq:pdf_single}, and denoting  $\psi^{(0)}$ as the digamma function  to get a simpler bound on the ergodic capacity
\begin{eqnarray}\label{eq:rate_single_bound}
	\bar{\eta}^{\rm SAR} \geq \frac{2{m^m}}{{{ \log}(2)}\phi^2S_0^{\phi^{2}} \gamma_0^{\frac{\phi^2}{2}} {{\left({2{\sigma ^2}} \right)}^{{\frac{\phi^{2}}{2}}}}} {\Gamma (m)} \sum \limits _{j = 0}^\infty \frac{{{K^j}{d_j}}}{[ \Gamma(j+1)]^2} \nonumber \\ \xi^{\frac{\phi^2}{2}}\Gamma(j+1)\big(-1-\frac{\phi^2}{2} {\log}(\xi)+ \frac{\phi^2}{2} \psi^{(0)} (j+1)   \big)
\end{eqnarray}
by applying integration-by-parts with the identity [\cite{Gradshteyn}, eq.4.352.1].

\subsection{Multi-antenna Reception with MRC}
\subsubsection{Outage Probability}
An exact expression of the outage probability for the MRC is $P_{\rm out}^{\rm MRC}= F_{\gamma}(\gamma_{\rm th})$, where $F_{\gamma}(\gamma_{})$ is given in \eqref{eq:cdf_mrc}. We use \cite{AboRahama_2018} to present the asymptotic outage probability in  \eqref{eq:asymptotic_out}. Considering the dominant terms at high SNR, we get the diversity order as $\sum_{l=1}^L \min\{j_l+1, \phi^2/2\}$. It is interesting to note that the diversity order is independent of FTR fading parameters $K$, $m$, and $\Delta$ as also observed in earlier literature \cite{Jiakang2019}, and has been extensively verified for THz transmissions in numerical section V. The diversity order shows the advantage of multi-antenna reception and  that the impact of pointing errors can be minimized by circumventing the multi-path fading using  a sufficiently higher beam-width.

\subsubsection{Average  BER}
	Substituting $F_{\gamma_{}}(\gamma)$  in  \eqref{eq:ber_eqn} and applying the definition of multivariate Fox’s H-function with the following  inner integral $I_1$
	\begin{eqnarray}
	\label{eq:ib}
	I_1\hspace{-1mm} = \hspace{-1.5mm}\int_{0}^{\infty} \hspace{-3mm}e^{-qz}z^{p-1}{z}^{\sum_{l=1}^L \left(\frac{\phi^2}{2} + \zeta_l\right)}dz\hspace{-0.5mm}=\hspace{-0.5mm}\frac{\Gamma\left(p+\frac{L\phi^2}{2}\sum_{l=1}^L\zeta_l\right)}{q^{p+\frac{L\phi^2}{2}+\sum_{l=1}^L\zeta_l}}
\end{eqnarray}
and applying the definition of multivariate Fox’s H-function on the resultant expression \cite{Kilbas_2004}, we get
\begin{eqnarray}\label{eq:ber_mrc}
	\bar{P}_e^{\rm MRC} =& \hspace{-1mm}\frac{1}{2\Gamma(p)q^{\frac{L\phi^2}{2}}}\sum \limits _{{[j_i=0]}_{i=1}^L}^\infty \hspace{-2mm}\prod_{l=1}^{L} \frac{{{\phi^{2}}{{m_l}^{m_l}}}}{{2S_0^{\phi^{2}}} {\Gamma (m_l)}} {{\frac{{{{K_l}^{j_l}}{{d_l}_{j_l}}}}{{[\Gamma^{}(j_l+1)]^2}{(2{\sigma_l^{2}}\gamma_0)^{\phi^2/2}}}}} \nonumber\\ &
	\hspace{-8mm}H^{0,1:2,1;2,1;\dots;2,1}_{2,1:2,2;2,2;\dots;2,2} \bigg[ ~\begin{matrix} F(\gamma_0) \end{matrix}~ \bigg| \begin{matrix} ~~F_1~~  \\ ~~F_2~~   \end{matrix}\bigg]
\end{eqnarray}
where $F(\gamma_0) = \big\{\frac{1}{{2\sigma_i^2 S_0^{2} \gamma_0 q}}\big\}_{i=1}^L$, $F_1 = \big\{ \big(1-p-\frac{L\phi^2}{2};1,\cdots,1 \big) \big\}: \big\{ \big(1-\frac{\phi^2}{2},1 \big), \big(1,1\big) \big\};\cdots;\big\{ \big(1-\frac{\phi^2}{2},1 \big), \big(1,1\big) \big\}$ and $F_2 = \big\{ \big(-\frac{L\phi^2}{2};1,\cdots,1 \big) \big\}; \big\{ \big(-\frac{\phi^2}{2}+j_i+1,1 \big), \big(0,1\big) \big\}_{i=1}^L$.  Similar to the outage probability, we get the asymptotic expression in \eqref{eq:asymptotic_ber} and  the diversity order as $\sum_{l=1}^L \min\{j_l+1, \phi^2/2\}$.

\begin{figure*}[t]
	\subfigure[Outage probability.]{\includegraphics[scale=0.28]{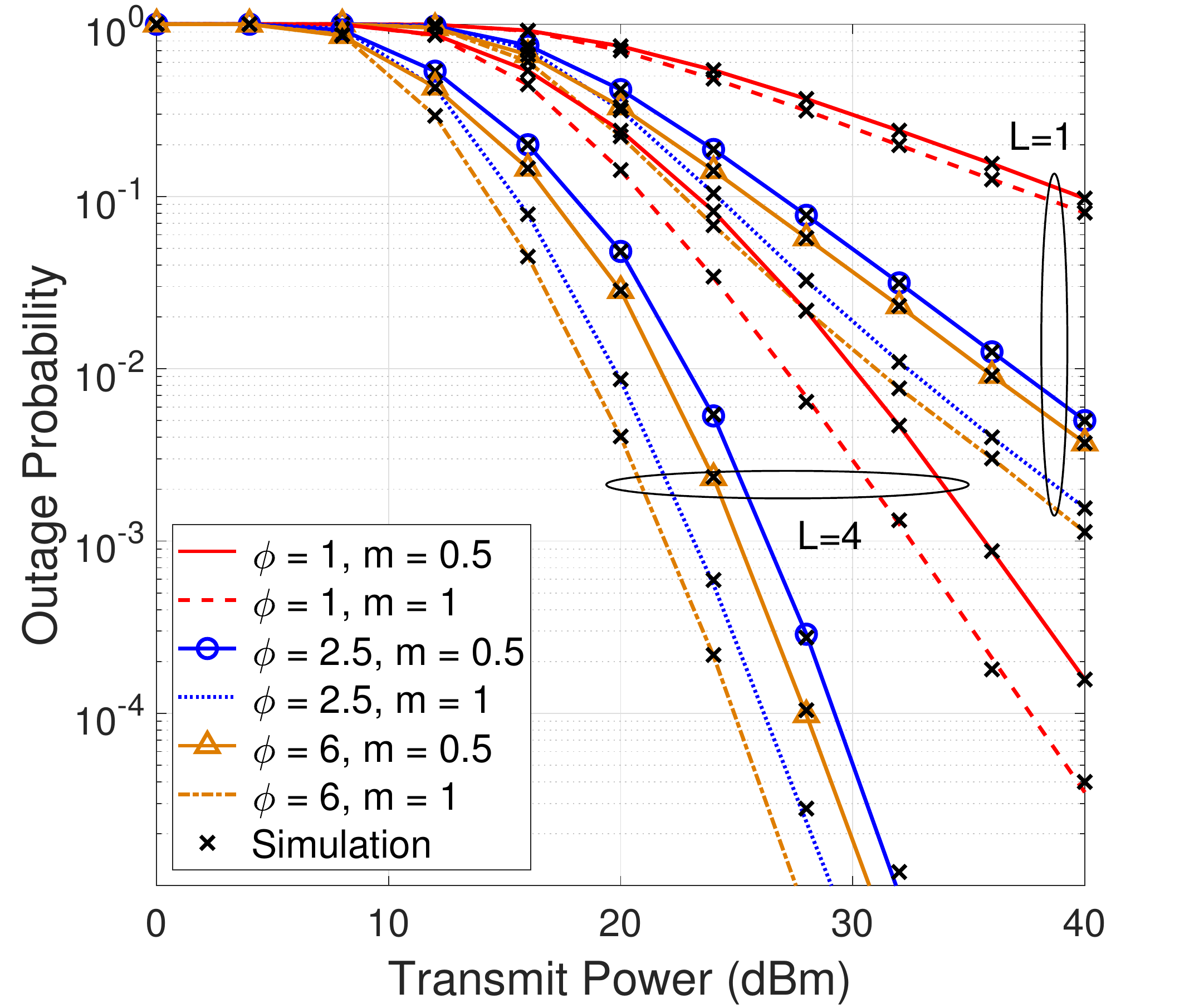}}
	\subfigure[Average BER.]{\includegraphics[scale=0.28]{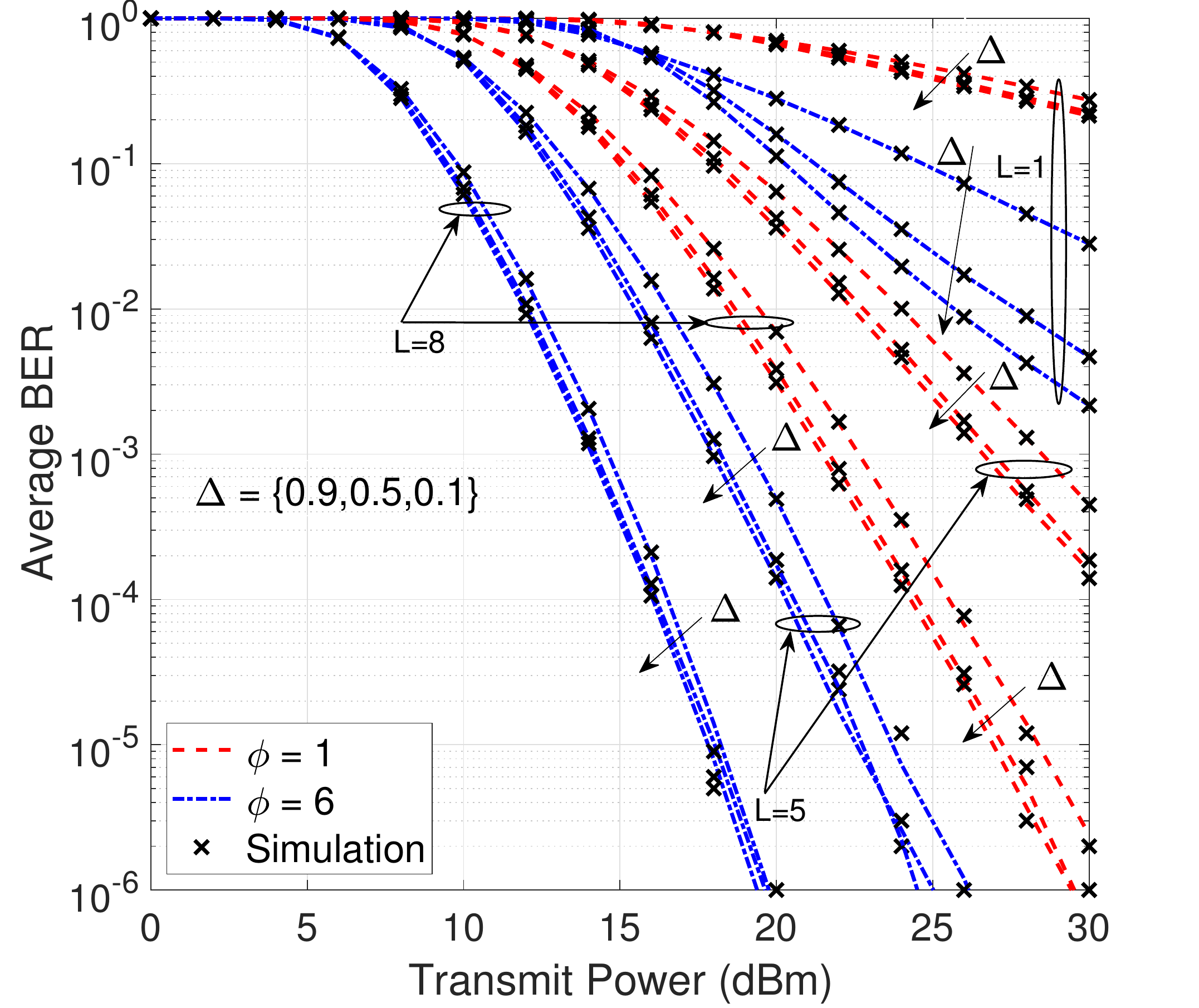}}
	\subfigure[Ergodic capacity.]{\includegraphics[scale=0.28]{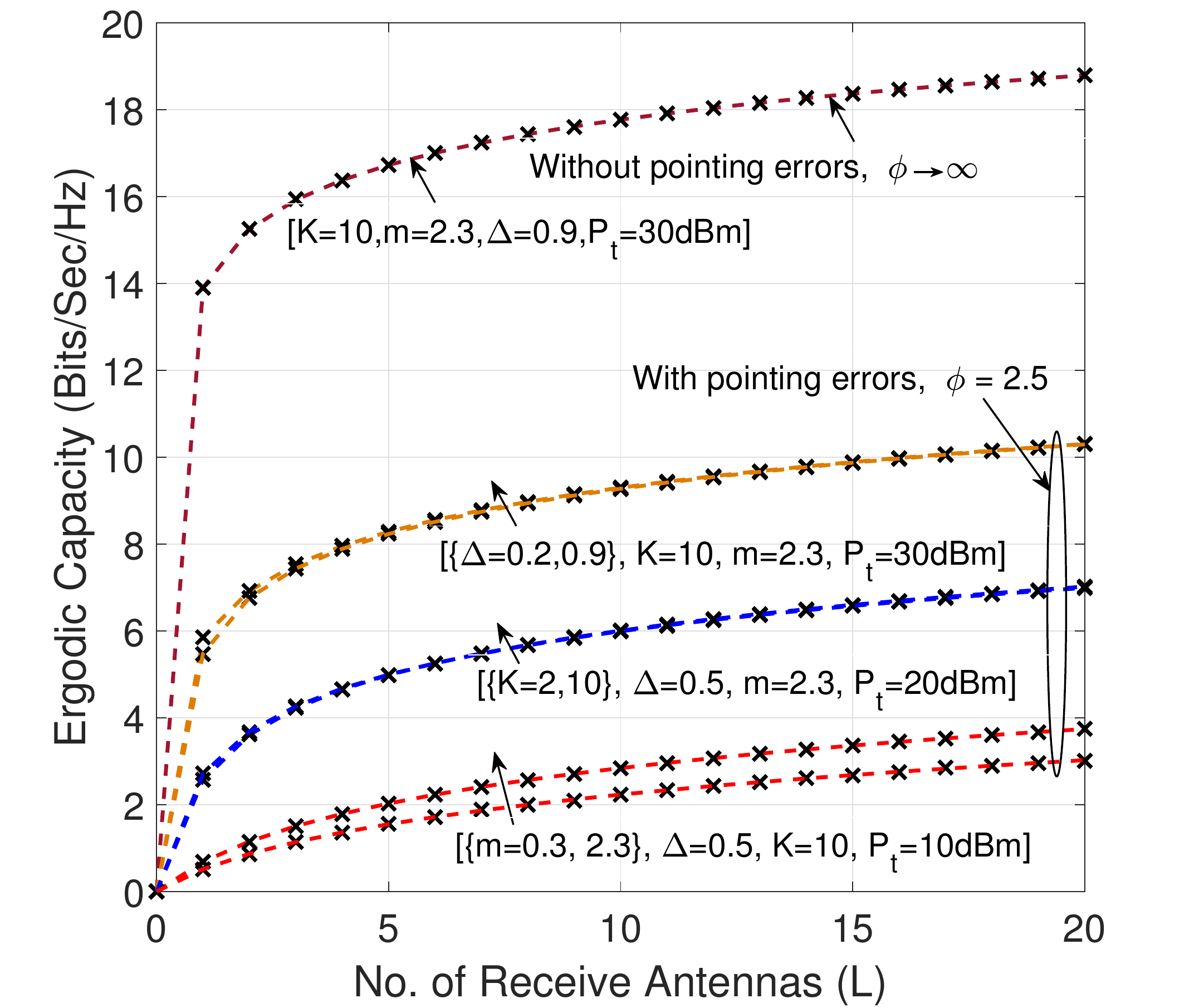}}
	\caption{Performance of THz wireless transmissions over FTR fading with pointing errors.}
	\label{fig:outage}
	\label{fig:ber}
	\label{fig:capacity_antenna}
\end{figure*}
\subsubsection{Ergodic Capacity}
Substituting $f_{\gamma}(\gamma)$  in \eqref{eq:rate_eqn} and applying the definition of multivariate Fox’s H-function \cite{Kilbas_2004}:
	\begin{eqnarray}
	\label{eq:cap_proof}
	&\bar{\eta}^{\rm MRC} = \hspace{-3mm}\sum \limits _{{[j_i=0]}_{i=1}^L}^\infty \hspace{-2mm}\prod_{l=1}^{L}\big( \frac{{{\gamma^{2}}{{m_l}^{m_l}}}}{{2A_0^{\gamma^{2}}} {\Gamma (m_l)}} {{\frac{{{{K_l}^{j_l}}{{d_l}_{j_l}}}}{{[\Gamma^{}(j_l+1)]^2}{(2{\sigma_l^{2}}\gamma_0)^{\gamma^2/2}}}}}\big)
	(\frac{1}{2\pi i})^L\nonumber \\ &\int_{\cal{L}}\frac{1}{\Gamma \big(\frac{L\gamma^2}{2} + \sum_{l=1}^L\zeta_l\big)} 
	\big\{\prod_{l=1}^{L}\frac{\Gamma\big(-\frac{\gamma^{2}}{2}+j_l+1 - \zeta_l\big)\Gamma(-\zeta_l)\Gamma\big(\frac{\gamma^{2}}{2}+\zeta_l\big)}{\Gamma\big(1-\zeta_l\big)}
	\nonumber \\&	\big(\frac{1}{{2\sigma_l^2 S_0^{2}}\gamma_0}\big)^{\zeta_l}d\zeta \big\}  \frac{1}{\ln(2)}\int_{0}^{\infty} \ln(1+z)
	{z}^{-1+\sum_{l=1}^L \left(\frac{\gamma^2}{2} + \zeta_l\right)} dz
\end{eqnarray}
We use the Mellin's inverse transform $ \ln(1+z)=\frac{1}{{2\pi i}}\int_{{L+1}} 
\frac{\Gamma\left(1+\zeta_{L+1}\right)\Gamma\left(-\zeta_{L+1}\right)\Gamma\left(-\zeta_{L+1}\right)}{\Gamma\left(1-\zeta_{L+1}\right)}{z}^{-\zeta_{L+1}}d\zeta_{L+1}$ to represent the inner integral in \eqref{eq:cap_proof} as
\begin{eqnarray}
	\label{eq:cap_ia_2}
	&	I_2 = \frac{1}{\ln(2){2\pi i}}\int_{L_{L+1}} 
	\frac{\Gamma\left(1+\zeta_{L+1}\right)\Gamma\left(-\zeta_{L+1}\right)\Gamma\left(-\zeta_{L+1}\right)}{\Gamma\left(1-\zeta_{L+1}\right)}  d\zeta_{L+1} \nonumber \\& \times \int_{0}^{\infty} {z}^{-1+\sum_{l=1}^L \left(\frac{\gamma^2}{2} + \zeta_l\right)} {z}^{-\zeta_{L+1}}dz
\end{eqnarray}
Since the inner integral in 	\eqref{eq:cap_ia_2} is not convergent, we use final value theorem $\lim_{t\rightarrow \infty} \int_{0}^{t}f(z) dz = \lim_{s\rightarrow 0} F(s) = F(\epsilon)$ with Laplace transform of  integrand in 	\eqref{eq:cap_ia_2}  to get
\begin{eqnarray}
	\label{eq:flt1}
	&	I_{21} = \int_{0}^{\infty} {z}^{-1+\sum_{l=1}^L \left(\frac{\gamma^2}{2} + \zeta_l\right)} {z}^{-\zeta_{L+1}}dz\nonumber \\&
	=\Gamma \left(\frac{L\gamma^2}{2} +\sum_{l=1}^{L+1}\zeta_l-\zeta_{L+1}\right) \left(\frac{1}{\epsilon}\right)^{\frac{L\gamma^2}{2} + \sum_{l=1}^{L+1}\zeta_l-\zeta_{L+1}}
\end{eqnarray}

Using \eqref{eq:cap_ia_2} and \eqref{eq:flt1} 	 in \eqref{eq:cap_proof},  and applying the definition of multivariate Fox’s H-function, we get 
\begin{flalign} \label{cap_lemma}
&\bar{\eta}^{\rm MRC}= 1.4427\sum \limits _{{[j_i=0]}_{i=1}^L}^\infty \prod_{l=1}^{L} \frac{{{\phi^{2}}{{m_l}^{m_l}}}}{{2A_0^{\phi^{2}}} {\Gamma (m_l)}}\nonumber \\ & {{\frac{{{{K_l}^{j_l}}{{d_l}_{j_l}}}}{{[\Gamma(j_l+1)]^2}{(2{\sigma_l^{2}}\gamma_0\epsilon)^{\phi^2/2}}}}} 
 H^{0,1:2,1;2,1;\dots;2,1;2,1}_{1,1:2,2;2,2;\dots;2,2;2,2}
\bigg[~ \begin{matrix} G(\gamma_0) \\ \epsilon \end{matrix} ~\bigg| \begin{matrix} ~~G_1~~ \\ ~~G_2~~ \end{matrix} \bigg]  
\end{flalign}
where $G(\gamma_0) = \big\{\frac{\gamma}{{2\sigma_i^2 S_0^{2} \gamma_0 \epsilon}}\big\}_{i=1}^L$, $G_1 = \big\{ \big(1-\frac{L\phi^2}{2};1,\cdots,1,-1 \big) \big\}; \big\{ \big(1-\frac{\phi^2}{2},1 \big), \big(1,1\big) \big\};\cdots;\big\{ \big(1-\frac{\phi^2}{2},1 \big), \big(1,1\big) \big\}; \big\{(0,1),(1,1)\big\} $ and $G_2 = \big\{ \big(1-\frac{L\phi^2}{2};1,\cdots,1,0 \big) \big\}; \big\{ \big(-\frac{\phi^2}{2}+j_1+1,1 \big), \big(0,1\big) \big\}_{i=1}^L; \big\{(0,1),(1,1)\big\}  $
where $\epsilon\to 0$.

\section{Simulation and Numerical Results}
In this section,  we demonstrate the performance of the considered single-antenna and MRC receivers for THz transmissions and validate the derived analytical expressions with  Monte Carlo simulations. We consider the THz channel with a distance of $50$\mbox{m}, carrier frequency $275$ \mbox{GHz}, and antenna gains of $50$ \mbox{dBi}. To compute the path loss for the THz link, we consider the relative humidity, atmospheric pressure, and temperature as $50\%$, $101325$ \mbox{Pa}, and $296$ \mbox{K}, respectively. We use \cite{Farid2007} to compute the  parameters $\phi$ and $S_0$ of pointing errors by varying the beamwidth and the jitter variance  $10$\mbox{cm}  antenna aperture  radius. The AWGN noise power is considered to be $-94.2$ \mbox{dBm}.

Fig. 1(a) demonstrates the impact of receiver antennas $L$, pointing errors parameter $\phi$, and FTR fading $m$ on the outage performance of the considered THz system with  $S_0=0.054$, $\gamma_{\rm th}=4$\mbox{dB}, $K=10$, and $\Delta=0.5$.  It can be seen that  improvement in the performance is significant with spatial diversity when   the number of receiver antennas is increased from $L=1$ to $L=4$. The figure also confirms  that the outage probability improves when the fading severity parameter $m$ increases, as expected. The slope of plots clearly demonstrate the dependence of diversity order on the system parameters.  It can be seen that there is a distinguishable  difference in the slopes  for $\phi = 1$ and $\phi = 2.5$ but there is a minimal change in the slope with $\phi = 2.5$ and $\phi = 6$.  As such, the diversity order increases linearly with $L$, is independent of the parameter $m$, depends on $\phi$ when $\phi^2/2< \min{j_l+1}$ but becomes independent otherwise. Thus, the impact of pointing errors can be mitigated using a sufficiently higher beam-width for THz transmissions.

We demonstrate the average BER performance by varying the parameters  $\Delta$ and $\phi$ with $K=2$, and  $m=2$.  The figure shows that highly  dissimilar specular components of FTR fading depicted through $\Delta$ provides an improved  average BER  performance. Similar to the outage probability, we can observe that the average BER improves with an increase in the number of receiver antennas $L$ and the diversity order is independent of $\Delta$ and becomes independent of pointing errors with a sufficiently higher value of  $\phi$.

Finally, Fig. 1(c) illustrates the relationship between the ergodic capacity and the number of receiver antennas with inter-dependence of the various channel and system parameters. The figure  shows the logarithmic scaling of the ergodic capacity with the number of receiver antennas $L$. It can be observed that  ergodic capacity is nearly independent of the parameters $K$ and $\Delta$,  but increases with an increase in the parameters $m$. The figure  also shows that pointing errors significantly degrade the THz performance. However,  an increase in the MRC antennas reduces the gap in performance by harnessing the spatial  as compared to the single-antenna system.

\section{Conclusions}
In this paper, we analyzed the performance of THz wireless transmissions under the  combined effects of path loss,  generalized FTR fading,  and Rayleigh distributed pointing errors. We provided statistical results on   the  single-antenna   and multi-antenna receivers by deriving PDF and CDF of the resultant SNR.  We analyzed the performance of the considered system by deriving closed-form expressions for the outage probability, average BER, and ergodic capacity. Using asymptotic analysis on the outage probability and average BER, we derived the diversity order of the system, which provides a design criterion of using sufficiently higher beam-width  to mitigate the impact of pointing errors by circumventing the multi-path fading. We validated our derived analytical expressions with Monte-Carlo simulations to show that the impact of the number of receiver antennas $L$,  pointing errors $\phi$,  and fading severity parameter $m$ is higher on the THz wireless system compared with other parameters. Incorporating hardware impairment in the performance analysis may be a possible extension of the proposed work.
 
\section*{Appendix A}
Using the definition of the MGF function, we apply inverse Laplace transform to find the PDF of $\gamma_{}= \sum_{i}^{L} \gamma_i$ as $f_{\gamma_{}}(\gamma)=  \mathcal{L}^{-1} M_{\gamma_{}}(s)$, where $M_{\gamma_{}}(s)=\prod_{i=1}^{L} M_{\gamma_{i}}(s)$ and $ M_{\gamma_{i}}(s)$ is the MGF of the $i$-th random variable $\gamma_i$.
Converting the incomplete Gamma function in \eqref{eq:pdf_single} to Meijer's G and applying the Meijer's G identity of definite integration, we get
\begin{eqnarray}
&M_{\gamma_{}}(s) = \prod_{l=1}^{L} \frac{{{\phi^{2}}{{m_l}^{m_l}}}}{{2S_0^{\phi^{2}}} {\Gamma (m_l)}} \sum \limits _{{[j_i=0]}_{i=1}^L}^\infty {{\frac{{{{K_l}^{j_l}}{{d_l}_{j_l}}{s^{-\frac{\phi^2}{2}}}}}{{[\Gamma^{}(j_l+1)]^2}{(2{\sigma_l^{2}{\gamma_0}})^{\phi^2/2}}}}}\nonumber \\&
G_{2,2}^{2,1}\left(\frac{1}{{2\sigma_l^2 S_0^{2} \gamma_0 s}}\left\vert \begin{matrix} {-\frac{\phi^2}{2} + 1,1} \\ {-\frac{\phi^{2}}{2}+j+1 , 0} \end{matrix} \right)\right.
\end{eqnarray}

Applying the definition of Meijer's G function \cite{Meijers} and interchanging the sum and product, we get
\small
\begin{eqnarray}
	\small
&M_{\gamma_{}}(s)= \sum \limits _{{[j_i=0]}_{i=1}^L}^\infty \prod_{l=1}^{L}\big( \frac{{{\phi^{2}}{{m_l}^{m_l}}}}{{2S_0^{\phi^{2}}} {\Gamma (m_l)}} {{\frac{{{{K_l}^{j_l}}{{d_l}_{j_l}}{s^{-\frac{\phi^2}{2}}}}}{{[\Gamma^{}(j_l+1)]^2}{(2{\sigma_l^{2}}{\gamma_0})^{\phi^2/2}}}}}\big)\nonumber \\&
\frac{1}{2\pi i}\int_{L_l}\frac{\Gamma\big(-\frac{\phi^{2}}{2}+j_l+1 - \zeta_l\big)\Gamma\left(-\zeta_l\right)\Gamma\big(\frac{\phi^{2}}{2}+\zeta_l\big)}{\Gamma\big(1-\zeta_l\big)}
\big(\frac{1}{{2\sigma_l^2 S_0^{2} \gamma_0 s}}\big)^{\zeta_l} d\zeta_l
\end{eqnarray}
\normalsize
Thus, using $f_{\gamma_{}}(z) = \frac{1}{2\pi i}\int_{L} e^{sz}  M_{\gamma_{}}(s) ds$, interchanging the integral, and rearranging the terms, we get
\begin{eqnarray}
\label{eq:pdf}
&f_{\gamma_{\rm }}(z) = \sum \limits _{{[j_i=0]}_{i=1}^L}^\infty \prod_{l=1}^{L}\big( \frac{{{\phi^{2}}{{m_l}^{m_l}}}}{{2S_0^{\phi^{2}}} {\Gamma (m_l)}} {{\frac{{{{K_l}^{j_l}}{{d_l}_{j_l}}}}{{[\Gamma^{}(j_l+1)]^2}{(2{\sigma_l^{2}\gamma_0})^{\phi^2/2}}}}}\big)\nonumber \\&
\prod_{l=1}^{L}
\frac{1}{2\pi i}\int_{L_l}\frac{\Gamma\big(-\frac{\phi^{2}}{2}+j_l+1 - \zeta_l\big)\Gamma(-\zeta_l)\Gamma\big(\frac{\phi^{2}}{2}+\zeta_l\big)}{\Gamma(1-\zeta_l)}
\nonumber \\&\big(\frac{1}{{2\sigma_l^2 S_0^{2} \gamma_0}}\big)^{\zeta_l}  \frac{1}{2\pi i}\int_{L} e^{sz} s^{-\sum_{l=1}^L \big(\frac{\phi^2}{2} + \zeta_l\big)}ds d\zeta_l
\end{eqnarray}
We substitute  $sz = -b $, and  apply the identity \cite{Gradshteyn} (eq. 8.315.1) to  solve the inner integral in \eqref{eq:pdf}:
\begin{equation}
\label{eq:I}
I = \frac{1}{2\pi i}\int_{L} e^{sz} s^{-\sum_{l=1}^L \left(\frac{\phi^2}{2} + \zeta_l\right)}ds=\frac{{z}^{-1+\sum_{l=1}^L \left(\frac{\phi^2}{2} + \zeta_l\right)}}{\Gamma \left(\sum_{l=1}^L\left(\frac{\phi^2}{2} + \zeta_l\right)\right)}
\end{equation}

We substitute \eqref{eq:I} in \eqref{eq:pdf} and apply the definition of multivariate Fox’s H function \cite{Kilbas_2004} to get \eqref{eq:pdf_mrc}. Finally, we use	$F_{\gamma_{}}(z) = \mathcal{L}^{-1} \prod_{i=1}^{N} \frac{M_{\gamma_i}(s)}{s}$ and apply the similar steps used in the derivation of   PDF to get the CDF in \eqref{eq:cdf_mrc}, which concludes the proof of Theorem  1.

\bibliographystyle{ieeetran}
\bibliography{Thz_references_others}
\end{document}